\documentclass[a4paper]{article}

 \usepackage{amsmath}
 \usepackage{graphicx}
 \usepackage{amssymb}
 \usepackage{amsfonts}
 \usepackage{latexsym}
 \usepackage{cite}
 \DeclareUnicodeCharacter{2212}{-}

\renewcommand{\baselinestretch}{1.1}
\renewcommand{\vec}[1]{{\bf #1}}

\def\beq{\begin{eqnarray}}
\def\eeq{\end{eqnarray}}

\begin{document}
\begin{center}
{\Large\bf Exact Foldy-Wouthuysen transformation for a Dirac equation describing the interaction of spin-1/2 relativistic particles with an external electromagnetic field}
\vskip 6mm

\textbf{Bruno Gon\c calves\footnote{
E-mail: bruno.goncalves@ifsudestemg.edu.br},
\
M\' ario M. Dias J\' unior\footnote{
E-mail: mariodiasjunior@ice.ufjf.br}
\ and \
Larissa F. Eleotério\footnote{
E-mail: larissaeleoterio@hotmail.com}
}
\vskip 4mm

$^{1}$Instituto Federal de Educa\c c\~ ao, Ci\^ encia e Tecnologia Sudeste de Minas Gerais \\
\centering{IF Sudeste MG, 36080-001, Juiz de Fora - MG, Brazil} \vspace{0.4cm}\\

$^{2}$ Departamento de F\' isica, ICE, Universidade Federal de Juiz de Fora,\\
UFJF, 36036-330, Juiz de Fora - MG, Brazil \vspace{0.4cm}\\

$^{3}$Universidade Federal de S\~ ao Jo\~ ao del-Rei,  \\
\centering{UFSJ, 36301-160, S\~ ao Jo\~ ao del-Rei - MG, Brazil} \vspace{0.4cm}\\

\end{center}

\begin{quotation}

\begin{abstract}
The Thomas-Bargmann-Michel-Telegdi (T-BMT) equation is derived using the Exact Foldy-Wouthuysen transformation. Extra new terms were found, and we discuss their possible physical applications. The main point of this work is to detail the procedure to get the general result. We explicitly present the choice of parametrization we used on the initial Hamiltonian and the motivations to take it. We emphasize that the final equations can depend on this choice, and it is possible to prevent the manipulations of the quadratic Hamiltonian become extremely cumbersome. More importantly, it is done in such a way that the transformed equations allow the direct separation into mass, kinetic, and interaction correction terms to the original T-BMT equation.    
\end{abstract}


\vspace{0.2cm}

\end{quotation}

\section{Introduction}

The Foldy-Wouthuysen transformation (FWT) was developed seventy years ago. It is a powerful technique to diagonalize the Dirac Hamiltonian performed step-by-step. At the end of the procedure, a non-relativistic approximate Hamiltonian is obtained with the equations to each bi-spinor decoupled. Already in the original paper \cite{foldy1950}, the authors showed some interesting applications of the method. A very fruitful approach to this technique is been 
used in many works since then. In general, the idea is to make explicit the possible interactions between external fields themselves and with the Dirac spinor on-shell. 

Some of the papers published just a few years after the original work are used as the basis for studies of the Dirac Hamiltonian with different external fields even nowadays. At \cite{foldy1952}, there is a description of the electromagnetic properties for Dirac particles moving in external electromagnetic fields. Two years later, in \cite{case1954}, there are some very useful generalizations of FW transformation for the spin-zero and spin-one theories. 

The works of Eriksen, which came just after, introduced a new method, base on the FW procedure that showed to be more general and intuitive. In \cite{eriksen1958}, the author demonstrates an exact solution with a generalization of FW transformation for the two-particle wave equations. But it was at \cite{kolsrud1960} that the exact transformation was presented. On one hand, the initial Hamiltonian has to satisfy some boundary conditions to allow the use of the method. On the other, once it is developed the transformed equations may have new information in comparison with the standard technique. 

There is a vast literature concerning these two approaches. The extension of the FW transformation and a modification of the method to obtain general results were presented at \cite{blount1962}. The development of the Dirac equation by the operator method was done at \cite{stephani1965}, in which the Hamilton operator is represented as a series in powers of $c^{-2}$. Two methods to obtain an approximate non-relativistic Hamiltonian for an electron in the electromagnetic field - Pauli's elimination method and FW transformation - were done in \cite{devries1968}. At \cite{suttorp1970} and \cite{groot1972}, the covariant equations of motion and spin are deduced for a charged particle with a magnetic dipole moment. There is also an exact solution with a generalization of FW transformation for the two-particle wave equations.  The book of \cite{berestetskii1982}, published in 1982, has a description of a method showing that the electromagnetic interactions of particles can be affected by a generalization of the method used in classical non-relativistic quantum theory. In \cite{golec1986}, the FWT was used o derive the relativistic equations of motion for a classical colored, spinning particle. In the '90s, there was a growing interest \cite{costella1995} in formalizing the method.  In \cite{nikitin1998} and \cite{violeta2000}, the general structure of the involution operator is presented with operators that act on the matrices and functions space. 

In 2001, a very interesting result was shown in \cite{obukhov2001}. The main point here was the deviation of the gravitational Darwin term through the EFW method. This was a non-expected term of the non-relativistic approximation of the Dirac particle coupled to the static spacetime metric since it did not come directly from the usual FWT. The natural interpretation here is to consider the EFWT a more general procedure in comparison with the canonical FWT. The exact derivation gives freedom in the choice of the parameter to perform the non-relativistic approximation of the final Hamiltonian. If the chosen coupling constant agrees with the method used to expand the exact solution, an unexpected term may become explicit at the final Dirac equation. This was applied in the works of  \cite{BGGW} and  \cite{BGtorsion}, where the main point was the search for an unknown term that could compensate the weakness of the gravitational waves, on the first, and gravitational torsion field, on the second, with an external strong magnetic field. A complete review on aspects of torsion can be found in \cite{buchshap} and \cite{shapiro2002}.

Prof. Silenko worked on many applications of this approach. In \cite{silenko2003}, the FW transformation for relativistic particles in electroweak fields. The study for relativistic particles with electric and magnetic dipole moments that interact with an electromagnetic field is shown in \cite{silenko2005}. He also showed in \cite{silenko2005semiclassical} that it is important to carefully take into account the choice of the expansion method to each case, since the Eriksen-Kolsrud transformation is not, in all cases, equal to the EFWT. In \cite{silenko2008}, the transformation for relativistic particles in a strong electromagnetic field was done and \cite{silenko2009} deals with the conditions of the transition from the Dirac representation to the FW one. There is a very interesting work, published in 2013, in which a complete comparison between the step-by-step method and the exact procedure is done in \cite{silenko2013comparative}. Also, the exact exponential operator of the FWT has been derived in \cite{silenko20163}. It is not just a good review of both techniques but an overview of the advantages and disadvantages of each approach. In addition, other works\cite{silenko2007,obukhov2009} show that the operators can present different aspects during the Hamiltonian diagonalization procedure.  

Recently, it was published a series of papers where the EFWT was generalized, giving the possibility to work with a more general set of Hamiltonians \cite{BMB2014}, \cite{BMBD2016} and \cite{FFP14}. In \cite{BMB2019}, an involution operator which allows many external fields with a big variety of coupling constants was explicitly shown. It guarantees the extension of the exact transformation to all Hamiltonians with analogous construction, as the initial one present in that paper. A thought-provoking discussion on results similar to those involving parity symmetry in the context of non-relativistic direct approximation can be located in \cite{BANERJEE2020114994}.

In this particular work, we aim to generalize the results presented in \cite{silenko1995}. The exact transformation is performed to the same action of that work. The work of 95 was developed using directly the non-relativistic approximation. We use the same notation of the original paper and the same final equations with all terms and the same coupling constants were found. Furthermore, some new terms in comparison with the original equations of motion are obtained. Since we also used the $1/m$ parameter to expand the transformed Hamiltonian, we believe that the new terms arrived for the same reason of the apparition of the gravitational Darwin in \cite{obukhov2001}.

The paper is organized as follows.  First, we give an overview of the main results of \cite{silenko1995}, emphasizing, the approach to reach each of the equations. Then the EFWT is derived showing the techniques used to maintain the same notations. The next step is the comparison of transformed equations of the two methods, and the new terms are made explicit. The conclusions are constructed with discussions about the possible interpretation of these new results.

\section{Starting with the non-relativistic approximation}

The main point of this work is the searching for new terms compared to the results present in \cite{silenko1995}. The initial Dirac-Pauli Hamiltonian for spin-$1/2$ particles interacting with an external electromagnetic field is

\begin{eqnarray}
H &=& {\vec{\alpha}}\cdot\left(\vec{p} - e\vec{A}\right) + \beta m + e\Phi 
+ \mu'\left(-\vec{\Pi}\cdot\vec{H} + i\,{\vec{\gamma}}\cdot\vec{E} \right)\: ,
\label{silenko95}
\end{eqnarray} 
where $\mu'$ is the anomalous magnetic moment \cite{vogel2009}, $\Phi$,$\vec{A}$ 
and $\vec{E}$,$\vec{H}$ are the potentials and strengths of the external 
electromagnetic field, ${\vec{\gamma}}$ are Dirac matrices 
and ${\vec{\sigma}}$ are Pauli matrices.
The operator $\vec{\Pi}$ describes the polarization of particles 
in this case and is written as
\begin{equation}
\vec{\Pi} = \left(\begin{array}{cc}
            {\vec{\sigma}} &  0\\
            0       & {\vec{-\sigma}} 
\end{array}\right) \; .
\label{Pi}
\end{equation}

To diagonalize the Hamiltonian in the original work, the author rewrote it as a relativistic expression in the FW representation. Then, the weak-field approximation was used, but the derivatives of all orders of the potentials were taken into account. This approach gave an equation of motion to the spin operator, from which it is possible to make the description of the anomalous magnetic moment in a sufficiently weak electromagnetic field for the Dirac particle.

The Dirac spinors can be written in two components $\phi$ and $\chi$ that are not independent, 
and one of these components can completely describe the interaction of the particle with an electromagnetic field. Thereby, the four-component Hamiltonian $H_{\Phi}$ is given by \cite{silenko1995}

\begin{eqnarray}
H_{\Phi} &=& \beta\varepsilon'\,+\,e\Phi\,+\,\frac{1}{4}\left\{\left(\frac{\mu_0m}{\varepsilon'+m}+\mu'\right)\frac{1}{\varepsilon'},\left(\Delta\Phi-2\vec{\Sigma}\cdot[\vec{E}\times\vec{p}]\right)\right\}_{+}\,-\,\cr
&-&\frac{1}{2}\beta\left\{\left(\frac{\mu_0m}{\varepsilon'}+\mu'\right),\vec{\Sigma}\cdot\vec{H}\right\}_{+}\,+\,\beta\frac{\mu'}{4}\left\{\frac{1}{\varepsilon'(\varepsilon'+m)},\left[(\vec{H}\cdot\vec{p})(\vec{\Sigma}\cdot\vec{p})\right.\right.\,+\,\cr
&+&\left.\left.(\vec{\Sigma}\cdot\vec{p})(\vec{p}\cdot\vec{H})+(\nabla\times \vec{H}\cdot \vec{p})+\frac{1}{2}(\vec{\Sigma}\cdot \nabla\times \nabla\times \vec{H})\right]\right\}_{+},
\label{finalsilenko}
\end{eqnarray}
where 
$\varepsilon'=\sqrt{{{\vec{\pi}}}^2 + {m}^2}$, 
${\vec{\pi}}=\vec{p}+e\vec{A}$, and 
$
\vec{\Sigma}=\left(\begin{array}{cc}
{\vec{\sigma}}   & 0 \\
0 & {\vec{\sigma}}      
\end{array}\right)
$.

The evolution of the polarization operator $\vec{\Pi}$ is described by the equation of spin motion. This equation for a bi-spinor is described by 

\begin{equation}
\dfrac{d\vec{\Pi}}{dt} \;=\; {i\Bigl[H_{\Phi}\, , \,\vec{\Pi}\Bigr]}\; .
\end{equation}
So, the equation of spin motion written for only one of the spinors is
\begin{eqnarray}
\dfrac{d{\vec{\sigma}}}{dt}&=&\frac{2}{\varepsilon}\left(\frac{\mu_0m}{\varepsilon+m}+\mu'\right)
\left[{\vec{\sigma}}\times\left[\vec{E}\times\vec{p}\right]\right]+2\left(\frac{\mu_0m}{\varepsilon}+\mu'\right)\left[{\vec{\sigma}}\times\vec{H}\right]\nonumber\\
&-&\frac{\mu'}{2\varepsilon(\varepsilon+m)}\Bigl\{(\vec{H}\cdot\vec{p})\left[{\vec{\sigma}}\times\vec{p}\right]+\left[{\vec{\sigma}}\times\vec{p}\right]\left(\vec{p}\cdot\vec{H}\right)+2\bigl[{\vec{\sigma}}\times\bigl[\left[\vec{p}\times\vec{H}\right]\times\vec{p}\bigr]\bigr]\Bigr\} .
\label{eqspinsilenko}
\end{eqnarray}
In this case, one shall make the change $\vec{\Pi}\rightarrow{\vec{\sigma}}\, , \,\vec{\Sigma}\rightarrow{\vec{\sigma}}$ \cite{silenko1995}.

In the next section, an analogous equation is derived. It is obtained from a completely different method. We use the EFWT as an approach to the initial problem of getting equation (\ref{eqspinsilenko}) without losing any term through the first-order approximations.

\section{Calculus with the Exact method from the beginning}

There is a step-by-step didactic description of the EFWT method in \cite{goncalves2009}, which we are following here to our specific case. First, we calculate the squared Hamiltonian ${\cal H}^{2}$. 

To simplify algebra, 
we adopt the following structure
\begin{equation}
{\cal H} \quad =\quad X\quad +\quad Y\quad +\quad W\quad +\quad Z \qquad 
\label{forma1}
\end{equation}
where
\begin{eqnarray}
X &=& {\vec{\alpha}}\cdot\left(\vec{p}-e\vec{A}\right)\nonumber\\
Y &=& \beta m \nonumber\\
W &=& e\phi \nonumber\\
Z &=& \mu'\left(-\vec{\Pi}\cdot\vec{H}+i\,{\vec{\gamma}}\cdot\vec{E}\right)
\label{quantidades1}
\end{eqnarray}

An advantage of the EFWT is the fact that it allows one to use some specific notations that make all the steps, 
during the calculations, very clear. So, in the end, the transformed equations have results that can be easily compared to those known from the literature. Moreover, 
the calculus of ${\cal H}^2$ can be performed in different approaches.
Equation (\ref{forma1}) presents a parametrized structure to facilitate the multiplication of its terms among themselves. The choice of separating the terms in the quantities shown in (\ref{quantidades1}) directly affects the final result of the EFWT and the physical interpretation of it. Here we are not neglecting any terms or taking any limits, we are just doing the procedure as intelligently as possible so that we have the best generalized Hamiltonian for this case. 

If we consider the case where a Dirac particle interacts with an external field such as the electromagnetic field, \cite{foldy1950,bjorken1965,schwabl2008} we can note that the Hamiltonian is written in a form that is separated into three distinct terms. The first term is proportional to $\beta$, the second is an even term 
$\varepsilon$, and the last is an odd term $\mathcal{O}$. The method is detailed at \cite{foldy1950,BMB2014,BMB2019,bjorken1965,schwabl2008}. So, if we also use a separation logic for the 
Hamiltonian (\ref{silenko95}), we can separate this equation in four different quantities (\ref{quantidades1}) to calculate the squared Hamiltonian algebraically. There are different possibilities for such separation of terms to occur. We can choose to perform direct multiplication with the initial Hamiltonian or separate this equation through more or fewer terms, for example. 

In the hypothesis of working with the direct multiplication of equation (\ref{silenko95}), we can see in the equation for the squared Hamiltonian, the lack of different terms that involve the external interaction with the electric and magnetic fields as well as with the potentials. In the case of separating the Hamiltonian into two quantities only as a term $X$ containing the external electric and magnetic fields and a term $Y$ with the other Hamiltonian parameters, it is intriguing that we continue to observe the lack of terms after the final multiplication. Furthermore, 
if we perform the process for three separate quantities as $X$, $Y$ and $Z$, for example, we get a similar result to those described above. In all these cases we continue to obtain a final result that important terms do not appear.

Then, we choose to separate and group the terms into four quantities based on the physical interpretation of each of them here. The first term $X$ is related to the kinetic momentum operator wherein $\vec{A}$ is the vector potential and $e$ is the charge of the particle. The second one $Y$ corresponds to the resting energy of the particle. The next term $W$ is the scalar potential of the electromagnetic 
field and the last one $Z$ corresponds to external electromagnetic fields.

The next step is to verify the anti-commutation relation between the Hamiltonian and the involution operator $J$. For this particular case, we use the operator presented in \cite{nikitin1998}. In other words, the electromagnetic scalar potential $\phi$ must be an odd function.

Taking into account the separation and grouping dynamics of terms in the Hamiltonian 
to calculate ${\cal H}^2$, we obtain
\begin{eqnarray}
{\cal H}^2 &=& (\vec{p}-e\vec{A})^2\;-\;e\vec{\Sigma}\cdot \vec{H} \;+\; m^2 \;+\; 2\beta em\phi\,+\, e^2\phi^2\,+\,(\mu')^2\vec{\Sigma}^2\cdot\vec{E}^2\,+\,
\nonumber\\
&+&
(\mu')^2\vec{\Sigma}^2\cdot\vec{H}^2\;+\;2(\mu')^2\gamma^5\vec{\Sigma}\cdot(\vec{E}\times\vec{H})\,+\,2e\,{\vec{\alpha}}\cdot(\vec{p}-e\vec{A})\phi\,+\, ie\,{\vec{\alpha}}\cdot\vec{E}\;-\;
\nonumber\\
&-&\,2\mu'm\vec{\Sigma}\cdot\vec{H}-2e\mu'\gamma^0(\vec{\Sigma}\cdot\vec{H})\phi \;-\;2e\mu'\gamma^0\vec{\Sigma}\cdot(\vec{E}\times\vec{A})\;-\;\mu'\gamma^0\Delta\phi\,+\,
\nonumber\\
&+&
2ie\mu'\gamma^0\gamma^5(\vec{\Sigma}\cdot\vec{E})\phi\;+\;
 2i\mu'\gamma^0\vec{E}\cdot\vec{p}\;-\;2\mu'\gamma^0\vec{\Sigma}\cdot(\vec{E}\times\vec{p})
\,+\,
\nonumber\\
&+&
i\mu'\gamma^0\vec{\Sigma}\cdot(\nabla\times\vec{E})\,-\,2ie\mu'\gamma^0\vec{E}\cdot\vec{A}\,+\,
2\mu'\gamma^0\gamma^5\vec{H}\cdot\vec{p}\,+\,
\nonumber\\
&+&
2i\mu'\gamma^0\gamma^5\vec{\Sigma}\cdot(\vec{H}\times\vec{p}) \,+\,
\mu'\gamma^0\gamma^5\vec{\Sigma} \cdot (\nabla \times \vec{H})\;-\; 
2e\mu'\gamma^0\gamma^5\vec{H} \cdot \vec{A} \: .
\label{h2}
\end{eqnarray} 

One thing to note about this result is that it contains some terms that are proportional to "i" and might suggest that the Hamiltonian derived from this quadratic form is not a Hermitian operator. However, these terms originate from the electric coupling constant in the initial Hamiltonian (\ref{silenko95}), which is well defined in \cite{groot1972} with this specific value. Another aspect to clarify is that we are not using the standard gamma matrices notation from \cite{bjorken1965}, but the one that is consistent with \cite{silenko1995}. In fact, \cite{silenko2021hermiticity} provides a detailed discussion on the self-adjointness and Hermiticity of Dirac Hamiltonians. It shows, for example, that the Laplace operator in cylindrical and spherical coordinates appears to be non-Hermitian, but it is actually self-adjoint, as long as these concepts\cite{reed1980methods} are rigorously defined.

Now, we assume that ${\cal H}^2$ can be written as follows
\begin{equation}
{\cal H}^2 \; = \; K^2 \; + \; Q  
\label{forma2}
\end{equation}
in which $K^2$ are terms related to ${{\vec{\pi}}}^2\,=\,{\left(\vec{p}-e\vec{A}\right)}^2$ 
and $m^2$, and $Q$ corresponds to terms that not have such relation. To take the squared root of

${\cal H}^2$ we consider that the following equation

\begin{equation}
{\cal H}^2 \;=\; K^2 \;+\; Q \;=\; K^2\left[1+\frac{Q}{K^2}\right] 
\label{forma3}
\end{equation}
can be written as

\begin{equation}
 \sqrt{{\cal H}^2} \;=\; \sqrt{K^2}\left[1+\frac{Q}{2K^2}\right] \;=\; 
 \sqrt{K^2} \;+\; \frac{Q}{2\sqrt{K^2}}\: .
 \label{forma4}
\end{equation}
Replacing,

\begin{equation}
\varepsilon' \;=\; \sqrt{{{\vec{\pi}}}^2+m^2} \;=\; \sqrt{K^2}\: ,
\label{varepsilon}
\end{equation}
Eq. (\ref{forma4}) can be written in the form
\begin{equation}
\sqrt{{\cal H}^2} \;=\; \varepsilon' \;+\; \frac{Q}{2\varepsilon'} \: .
\end{equation}

After calculating the square root and some algebra, the transformed Hamiltonian 
can be presented as
\begin{eqnarray}
{\cal H}^{tr} &=& \beta\varepsilon'\,-\,e\phi\,+\,\frac{ie}{\varepsilon'} \vec{\Pi}\cdot(\vec{p}-e\vec{A})\phi\,+\,\beta\frac{e^2}{2\varepsilon'}\phi^2\,+\,\frac{\mu'}{2\varepsilon'}\Delta\phi \,+\,\frac{i\mu'}{\varepsilon'}\vec{E}\cdot(\vec{p}-e\vec{A})\,+
\nonumber\\
&+&
\beta\left(\frac{\mu_0m}{\varepsilon'}+\mu'\right)(\vec{\Sigma}\cdot\vec{H})\,-\,
\frac{e\mu'}{\varepsilon'}(\vec{\Sigma}\cdot\vec{H})\phi\,-\,\frac{e\mu'}{\varepsilon'}(\vec{\Sigma}\cdot\vec{E})\phi \,+\,
\mu_0(\vec{\Pi}\cdot\vec{E}) \,-
\nonumber\\
&-&
\frac{\mu'}{\varepsilon'}\vec{\Sigma}\cdot(\vec{H}\times\vec{p})\,-\,
\frac{\mu'}{\varepsilon'}\vec{\Sigma}\cdot\bigl[\vec{E}\times(\vec{p}-e\vec{A})\bigr]\,-\,\beta\frac{i(\mu')^2}{\varepsilon'}\vec{\Sigma}\cdot(\vec{E}\times\vec{H})\,+
\nonumber\\
&+&
\frac{i\mu'}{\varepsilon'}\vec{H}\cdot(\vec{p}-e\vec{A})\;+\;\frac{i\mu'}{2\varepsilon'}\vec{\Sigma}\cdot(\nabla\times\vec{H})\,+\,\frac{i\mu'}{2\varepsilon'}\vec{\Sigma}\cdot(\nabla\times\vec{E})+\;
\nonumber\\
&+&
\beta\frac{(\mu')^2}{2\varepsilon'}\bigl({\vec{H}\bigr)}^2
\;+\;\beta\frac{(\mu')^2}{2\varepsilon'}\bigl({\vec{E}}\bigr)^2 \; .
\label{finalBML}
\end{eqnarray}

Then let us consider the two-component spinor in the following form
\begin{equation}
{\Psi}_{\Phi} \,=\, \left(
\begin{array}{ccc}
\psi \\
\zeta  \\
\end{array}
\right) \,\exp^{-imt}\,.
\end{equation}
and write the Dirac equation in the Schr\" odinger form $i\partial_{t}\psi={\cal H}\psi$. 
After some algebra and taking such considerations into account, one can get the Hamiltonian 
to $\psi$ as
\begin{eqnarray}
H_{\psi}&=& m \,+\, \frac{1}{2m}\left\{(1+{A})\bigl[(\delta^{ij}+{B}_{ij}){P^{i}} \,+\, {C_{j}}\bigr]^2+D\right\} \: ,
\label{forma5}
\end{eqnarray}
where
\begin{eqnarray}
{A}&=& \frac{2(\mu_0)^2}{e}\;\sigma^{m}H_{m}\: . \nonumber\\
{B_{ij}}&=& \frac{\mu'}{2m}\Bigl[H_{i}\sigma_{j}\;+\;\sigma_{i}H_{j}\Bigr] \: . \nonumber\\
{C_{j}}&=& -\,2\mu'\varepsilon_{jkl}\sigma^{k}E^{l}\,-\,\frac{\mu'}{2m}\varepsilon_{jik}\partial^{i}H^{k}\,-\,2\mu'\varepsilon_{jkl}\sigma^{k}H^{l}\,+\,2i\mu'E_j\,+\,2i\mu'H_j\,+\,
\nonumber\\
&+&
2ie\gamma^0\sigma_j\phi \: . \nonumber\\
D&=& 2me\phi\,-\,2m(\mu_0+\mu')\sigma^{m}H_{m}\,+\,\mu'\Delta\phi\,+\,\frac{1}{2}\varepsilon_{knt}\varepsilon^{tms}\sigma^{k}\partial^{n}\partial_{m}H_{s}\,+\,e^2\phi^2\,-\, \nonumber\\
&-&2e\mu'\gamma^0\sigma^{j}E_{j}\phi\,-\,2e\mu'\gamma^0\sigma^{j}H_{j}\phi\,+\,2m\gamma^0\mu_0\sigma^{m}E_{m}\,+\,i\mu'\varepsilon_{kil}\sigma^{k}\partial^{i}E^{l}\,+\, \nonumber\\
&+&i\mu'\varepsilon_{kim}\sigma^{k}\partial^{i}H^{m}\,-\,2i(\mu')^2\varepsilon_{klm}\sigma^{k}E^{m}H^{l}\,+\,(\mu')^2E^{j}E_{j}\,+\,(\mu')^2H^{i}H_{i} \: .
\label{forma6}
\end{eqnarray}

Equation (\ref{forma5}) has a known didactic structure \cite{BMB2019}. These notations are used to simplify algebra and facilitate the interpretation of results. 
According to this equation, the first term corresponds to rest energy, and 
the second represents the kinetic term which is of the kind  
$\left(\vec{P}\,-\,e\vec{A}\right)$. The quantity $C_{j}^{tr}$ can be 
imagined as an analogous term of a gauge transformation to $P^i$ (in the 
situation where $B_{ij}^{tr}=0$). The quantity $(1\,+\,A^{tr})$ can be seen as a correction to the general form of the kinetic energy. Already the last term $D^{tr}$ corresponds to external interaction.

To quantize the Hamiltonian (\ref{forma5}) and write the equation 
of motion for the spin, wherein

\begin{equation}
\dfrac{d{\vec{\sigma}}}{dt} \;=\; {i\Bigl[H_{\psi}\, , \,{\vec{\sigma}}\Bigr]}\; .
\label{eqspin}
\end{equation}
So we get
\begin{eqnarray}
\dfrac{d{\vec{\sigma}}}{dt}&=&\,+\,\frac{2\mu'}{m}\bigl[{\vec{\sigma}}\times\bigl[\vec{E}\times\bigl(\vec{p}-e\vec{A}\bigr)\bigr]\bigr]\,+\,2(\mu_0+\mu')\bigl[{\vec{\sigma}}\times\vec{H}\bigr]\,-\, 
\nonumber\\
&-&\frac{\mu'}{4m^2}\Bigl\{(\vec{H}\cdot\vec{p})\bigl[{\vec{\sigma}}\times\vec{p}\bigr]\;+\;\bigl[{\vec{\sigma}}\times\vec{p}\bigr](\vec{p}\cdot\vec{H})\;+\;2\bigl[{\vec{\sigma}}\times\bigl[[\vec{p}\times\vec{H}]\times\vec{p}\bigr]\bigr]\Bigr\}\,-\, 
\nonumber\\
&-&\frac{2(\mu_0)^2}{em}\bigl(\vec{p}-e\vec{A}\bigr)^2\bigl[{\vec{\sigma}}\times\vec{H}\bigr] +\frac{2\mu'}{m}\bigl[{\vec{\sigma}}\times\bigl(\vec{H}\times\vec{p}\bigr)\bigr] -\frac{2ie}{m}\bigl[{\vec{\sigma}}\times\bigl(\vec{p}-e\vec{A}\bigr)\bigr]\phi +
\nonumber\\
&+&
\frac{2e\mu'}{m}\bigl[{\vec{\sigma}}\times\vec{H}\bigr]\phi\,+\,\frac{2e\mu'}{m}\bigl[{\vec{\sigma}}\times\vec{E}\bigr]\phi\,-\, \frac{i\mu'}{m}\bigl[{\vec{\sigma}}\times\bigl(\nabla\times\vec{H}\bigr)\bigr]\,-\,
\nonumber\\
&-&
\frac{i\mu'}{m}\bigl[{\vec{\sigma}}\times\bigl(\nabla\times\vec{E}\bigr)\bigr]
\,-\,2\mu_0\bigl[{\vec{\sigma}}\times\vec{E}\bigr]\,+\,\frac{2i(\mu')^2}{m}\bigl[{\vec{\sigma}}\times\bigl(\vec{H}\times\vec{E}\bigr)\bigr]\, .
\label{eqspinBML}
\end{eqnarray}

Next section deals with the interpretation of the new terms in (\ref{eqspinBML}) and its possible physical applications.

\section{The new terms}
\label{THT}

It is possible to see that there are thirteen new terms in (\ref{forma6}) when 
compared to (\ref{finalsilenko}). The last four terms in the quantity ${C}_j$ are new. 
They are related to the distributive property between the quantities $X$ and $Z$ for the
first three terms and between $X$ and $Y$ for the last one. Now, the other new terms are 
present in quantity $D$. The fifth term is the squared quantity $W$; the sixth and seventh terms are related to the operation between 
the quantities $W$ and $Z$; the eighth term with the quantities $X$ and $W$; the ninth and tenth terms are related to the quantities $X$ and $Z$; the eleventh, twelfth and thirteenth terms are the result of the squared quantity $Z$.

Whereas when we compare the equations of motion for the spin operator (\ref{eqspinsilenko}) 
and (\ref{eqspinBML}), one can see that eight new terms are 
presented last in equation (\ref{eqspinBML}). The first two terms come from 
the quantity ${C}_j$, the distributive property between the quantities $X$ and $Z$, 
and between $X$ and $W$, respectively. The others come from the quantity $D$. 
The third and fourth terms are related to the operation between the quantities 
$W$ and $Z$; the fifth and sixth terms are related to the quantities $X$ and $Z$; 
the seventh term is related to the quantities $X$ and $Y$; and the eighth, 
the squared quantity $Z$.

The next step should be the careful interpretation of each one of the thirteen extra 
terms of (\ref{eqspinBML}). Some derivations have been presented recently considering 
the generalization of the Bargmann-Michel-Telegdi (BMT) equation based on different 
interpretations, see \cite{khriplovich1998,rebilas2011,fukuyama2013,silenko2015spin}. 
The equations of motion for spin can also be presented in different ways: considering 
the Wentzel-Kramers-Brillouin approximation to the Dirac equation \cite{rubinow1963,rafanelli1964}, the heuristic arguments of \cite{berestetskii1982,khriplovich1998,fukuyama2013} and 
with the FW transformation, described at \cite{silenko2005}. An extensive 
number of works have been dedicated to study this topic since 
the BMT equation was published in \cite{bargmann1959} and 
the dynamics of spin in \cite{costella1994,pomeransky1999,mane2005,silenko2015edm}. 

The relativistic dynamics of electrons can be described by different theories applied 
in several areas of physics \cite{silenko2008,chen2014,frenkel1926} and the spin 
effects appear in condensed matter \cite{shen2005}, 
in quantum plasma \cite{marklund2007,brodin2007}, 
in astrophysical systems \cite{mahajan2014}, in gravitational field \cite{obukhov2009}, 
and in electromagnetic fields. \cite{hu1999,walser2002,roman2003,faisal2004,klaiber2014,zimmermann2015}

However, some works have shown that the FW representation is in agreement with 
Dirac theory \cite{dirac1928} than the Frenkel representation \cite{wen2016}, 
and the EFWT presents more information \cite{obukhov2001} than 
the usual transformation as mentioned before. 

In this case, one possibility to explain and understand the motion of 
the particle comes from ${\vec{\sigma}}\cdot(\vec{E}\times\vec{p})$, 
which is the spin-orbit coupling term in Hamiltonian (\ref{finalsilenko}) 
and (\ref{finalBML}), and is related to the direction of the 
propagation–polarization plane \cite{hu1999,walser2002,roman2003,klaiber2014}. 
In addition, another effect that can be noticed and analyzed is the possibility 
of the spin current appearing as a quantum response \cite{murakami2003,engel2005,pallab2011}, 
for example, through the term ${\vec{\sigma}}\times\vec{E}$ 
in equation (\ref{eqspinBML}).

\section{Discussions and conclusions}
\label{con}

We have considered Eq. (\ref{silenko95}) as the starting point to obtain the transformed Hamiltonian (\ref{forma6}). However, we wrote the initial Hamiltonianin the form presented in (\ref{forma1}) and (\ref{quantidades1}) to facilitate the calculations and interpret the results obtained with the EFWT. It is important to emphasize the fact that the EFWT is not equivalent to the FWT as it was explained above: the exact procedure should present some additional terms when compared to the usual one. The new terms were found, and the equations of motion to the spin operator were also compared as well. 

It was also shown that, if the initial Hamiltonian is written using four 
different parameters (\ref{quantidades1}), the result can be condensate form divided into kinetic and interaction terms (\ref{forma5}), with explicit corrections to each of them. It is important to note that the general Hamiltonian (\ref{h2}) may also give new information, once its expansion procedure is generalized. In this sense, it is possible to imagine the search for the correct parametrization steps that would give the known physical empiric interaction term derived from the initial action of the theory. 

The analysis of the new terms present in the non-relativistic equation of motion of the spin operator may lead to an easier understanding of their physical interpretation since there are many experiments related to the electromagnetic field-spin interaction. We treated some phenomenological aspects here, but we believe that a deeper search in the literature to each of these terms would give a richer view to each of them. 

The extra interactions shown in (\ref{eqspinBML}) related to (\ref{eqspinsilenko}) may be seen as a generalization of the T-BMT equation \cite{thomas1927,bargmann1959}, which describes the classical spin dynamics of a particle with anomalous magnetic moment and electric dipole moment in the electromagnetic field.

\section*{Acknowledgments}
\label{Ack}

The authors wish to thank Prof. Ilya L. Shapiro 
for the discussions about the problem. BG and LFE are 
grateful to Funda\c c\~ ao Nacional de Desenvolvimento da 
Educa\c c\~ ao (FNDE) for financial support. MDJ is grateful 
to the Programa de Bolsas de P\' os-Gradua\c{c}\~ ao da UFJF 
(PBPG-UFJF).

\renewcommand{\baselinestretch}{0.9}
\bibliography{biblio.bib}
\bibliographystyle{unsrt}

\end{document}